\documentclass{SCIS2017}
\usepackage{changepage}
\usepackage{url}

\begin{document}

\ArticleType{RESEARCH PAPER}
\Year{2017}
\Month{January}
\Vol{60}
\No{1}
\DOI{}
\ArtNo{}
\ReceiveDate{}
\AcceptDate{}
\OnlineDate{}

\title{CHAOS: an SDN-based Moving Target Defense System}{CHAOS: an SDN-based Moving Target Defense System}

\author[1]{Juan Wang}{{jwang@whu.edu.cn}}
\author[1]{Feng Xiao}{}
\author[1]{Jianwei Huang}{}
\author[1]{Daochen Zha}{}
\author[2]{Hongxin Hu}{}
\author[1]{Huanguo Zhang}{}

\AuthorMark{Juan Wang}

\AuthorCitation{Juan Wang, Feng Xiao, Jianwei Huang, et al}


\address[1]{Computer School, Wuhan University, Wuhan {\rm 430072}, China}
\address[2]{Division of Computer Science, School of Computing, Clemson University, Clemson {\rm SC 29634}, America}

\abstract{The static nature of current cyber systems has made them easy to be attacked and compromised. By constantly changing a system, Moving Target Defense (MTD) has provided a promising way to reduce or move the attack surface that is available for exploitation by an adversary. However, the current network-based MTD obfuscates networks indiscriminately that makes some networks key services, such as web and DNS services,  unavailable, because many information of these services has to be opened to the outside and remain real without compromising their usability. Moreover, the indiscriminate obfuscation also severely reduces the performance of networks. In this paper, we propose CHAOS, an SDN (Software-defined networking)-based MTD system, which discriminately obfuscates hosts with different security levels in a network. In CHAOS, we introduce a Chaos Tower Obfuscation (CTO) method, which uses a Chaos Tower Structure (CTS) to depict the hierarchy of all the hosts in an intranet and provides a more unpredictable and flexible obfuscation method.  We also present the design of CHAOS, which leverages SDN features to obfuscate the attack surface including IP obfuscation, ports obfuscation, and fingerprint obfuscation thereby enhancing the unpredictability of the networking environment. We develop fast CTO algorithms to achieve a different degree of obfuscation for the hosts in each layer. Our experimental results show that a network protected by CHAOS is capable of decreasing the percentage of information disclosure effectively to guarantee the normal flow of traffic.}


\maketitle


\section{Introduction}

Nowadays, the network security issues become increasingly prominent as all kinds of network security events emerge one after another. However the tranditonal network security tools cannot effectively defend increasingly complex and intelligent penetration of network intrusion and unknown vulnerability attacks. As usually, adversaries can break through or bypass firewalls and IDSes so that an intranet can be easily compromised. As one of revolutionary technologies, MTD  (Moving Target Defense) changes game rules, providing a dynamic and proactive network defense\cite{1}\cite{31}\cite{33}.

MTD aims at building a dynamically and continually shifting and changing system to increase complexity and cost for attackers, limit the exposure of vulnerabilities and opportunities for attackers, and increase system resiliency \cite{11}. The idea of MTD has been applied to network security, e.g. DYNAT \cite{12} and DESIR\cite{30}. The difference between MTD and traditional network tools, such as firewall and IDS, is that the latter will suspend suspicious actions once they break security rules. That lets adversaries to easily figure out the deployed network defense mechanism so that they will try to bypass them. However, MTD sends illegible fake information to potential threateners to make them spend more time and cost so that they will leave more footprints, making them easier to be exposed.

However,due to its closed and static characteristics, traditional network is difficult to realize dynamic and active security defense effectively and comprehensively. As a new type of network security architecture, SDN points a brand-new path for building dynamic and proactive defense system. SDN has a couple of benefits. It decouples network control and data planes, enabling network control to become directly programmable. It enables network managers to configure, manage, secure, and optimize network resources very quickly via dynamic and automated SDN programs. Meanwhile, SDN lets the underlying infrastructure to be abstracted from applications and network services. In addition, SDN controllers can provide a global view of the network. The central management of SDN makes networks more intelligent. 

Therefore, our goal is to build an SDN-based dynamic network defense system. In order to realize the SDN-based MTD, it has some key challeges to resolve. Firtly, we should leverage SDN to obfuscate network fingerprinting. Secondly, the moving target defense may make some networks services unavailable, such as DB server. Since the IP address and port number of these services have to be opened to the outside and remain real. If MTD obfuscates these services fully, it will return users with fake IPs and ports, making these services unable to use. Thirdly, obfuscating network parameters indiscriminately will severely reduce the performance of networks undoubtedly.

Motivated by the aforementioned goals and challenges, we propose CHAOS, a SDN-based MTD system. Utilizing the programmability and flexibility of SDN, CHAOS obfuscates the attack surface including IP obfuscation, ports obfuscation, and fingerprint obfuscation thereby enhancing the unpredictability of the networking environment. Furthermore it discriminately obfuscates hosts with different security levels in networks. In CHAOS, we propose the Chaos Tower Obfuscation (CTO) method, which uses the Chaos Tower Structure (CTS) to depict the hierarchy of all the hosts in an intranet and define expected connection and unexpected connection. Moreover, we develop fast CTO algorithms to achieve a different degree of obfuscation for the hosts in each layer. We design and implements CHAOS as an application of SDN contoller. Our approach lets it very easy to realize moving target defense in networks.

Furthermore, we evaluate our system and the results show that CHAOS can effectively hide real information of the target hosts from attackers as well as produce fake responses, which can disrupt an adversary's ability to sniff network traffic effectively. In addition, our tests show that the system have lower cost when compared to an unconditionally obfuscating system, which strengthens its applicability in real networks.

\begin{adjustwidth}{0.5cm}{0cm}
Our contributions can be summarized as follows:
\end{adjustwidth}

\begin{adjustwidth}{0.5cm}{0cm}
\begin{itemize}
\item We propose a new SDN-based MTD approach, CHAOS, where a Chaos Tower Structure (CTS) is constructed to represent a hierarchy of all the hosts on the network. Using the CTS, we can determine if a network connection is needed to be obfuscated. 
\item We present a more unpredictable and flexible obfuscation method named Chaos Tower Obfuscation (CTO) in CHAOS, where the level of obfuscation is decided reasonably. This method is expected to provide a good protection for hosts with relatively higher privileges but it will not interfere with normal communications. Furthermore, by sending instructions from SDN controllers, CHAOS can flexibly forward, modify all packets in network to obfuscate the attack surface including IP obfuscation, ports obfuscation, and fingerprint obfuscation. 
\item We design and implement CHAOS as an SDN application and evaluate its performance. The results demonstrate that a network protected by CHAOS can decrease the percentage of information disclosure effectively and has a lower cost. Thus, CHAOS is practical and can be used in the real-world systems instead of a theory model.
\end{itemize}
\end{adjustwidth}

The remainder of this paper is organized as follows. Section 2 provides some background information relating to our system. Section 3 describes how we design our system. Section 4 shows the details of the Chaos Tower Obfuscation (CTO) method. Section 5 presents the implementation and evaluation of our system. Section 6 shows some related work. Section 7 concludes this paper.

section{Background and Threat Model}
In this section, we first provide an introduction to SDN and its mechanism of asynchronous messaging. Then we introduce a threat model about our system.

\subsection{SDN and its Asynchronous Messaging Mechanism}
Software-defined networking (SDN) has emerged as a programmable and centrally controlling architecture providing an agile platform for vendors as well as enterprise users to control and define network.

The SDN controller plays the role of an operating system (OS) for networks \cite{2}. All communications between network applications and network devices have to go through the controller. OpenFlow protocol as the first SDN standards, defined the communication protocol between the SDN Controller and the forwarding plane of network devices such as switches and routers. The controller uses the OpenFlow protocol to control network devices and choose the best path for application traffic. Because the network control plane can be programmed, contrary to the firmware of hardware devices, network traffic can be managed more dynamically and at a much more granular level.  In other words, the whole network is controlled by the controller and its stability is also guaranteed by the controller.

Centralized control allows the SDN core controller to define the data flows \cite{1}. Each flow through the network must first get permission from the controller, which verifies that the communication is permissible by the network policy \cite{4}.  

\textbf{Flow table.} The OpenFlow switch contains the flow tables, which are used to perform packet lookups and forwarding \cite{4}. Using OpenFlow protocol, the controller can add, update, and delete flow entries in the flow table, both reactively (in response to packets) and proactively \cite{4}. Each flow table in the switch contains a set of flow entries. Each flow entry consists of matching fields, counters, and a set of instructions to apply to matching packets \cite{1}.If a packet matches the fields defined in the flow table, the instructions (i.e., ``actions'') are executed. If no match is found, a packet may be forwarded to the controller or continue to the next flow table.

\textbf{Packet-in message.} For all packets that do not have a matching flow entry, a packet-in event may be sent to the controller. There are mainly two situations that produce these messages: A mismatch in the tables of the switch or a time to live (TTL) error \cite{5}. Packet-in messages contain a variety of information about the flow.

After receiving the packet-in message, the core controller decides how to process irregular flows by dispatching a packet-out message.

\textbf{Packet-out message.} Packet-out messages are sent from the controller to a switch when the controller wishes to instruct the switch to send packets via a specified port of the switch, or to instruct the switch how to forward packets received via packet-in messages.

\subsection{Threat Model}
In most cases, adversaries start an attack to a intranet by collecting as much information about the network as they can. Then they connect to those vulnerable hosts and send attack payloads. Our system, CHAOS, aims to build a dynamic and variable network, so as to defeat reconnaissance attacks for an intranet. Thus, we assume an adversary can scan a network and monitor the network traffic. Moreover, the adversary can eavesdrop network packets. We also assume the protected networks is able to support OpenFlow-based SDN switches and controllers.

\begin{figure*}
\centering
\includegraphics[width=4.5in]{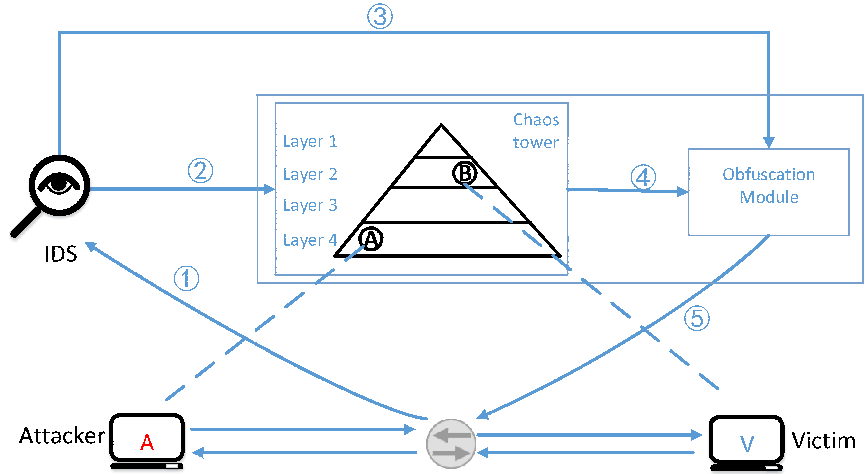}
\caption{Overall system of CHAOS. \textcircled{1} shows that IDS is monitoring the flows of OF switch; \textcircled{2}  and \textcircled{4} represent the process how CHAOS deals with the suspicious connections detected by IDS; \textcircled{3} means the normal connections determined by IDS will be allowed directly; \textcircled{5} means that the OF switch is controlled by CHAOS.}
\end{figure*}

\section{CHAOS Design}

In this section, we provide an overview of CHAOS and then highlight the design of Chaos Tower Structure (CTS).

\subsection{CHAOS System Overview}
The overall system is illustrated in Figure 1. We design two main modules: Chaos Tower Structure (CTS) and Obfuscation module (CTO). CTS defines the communication rules of hosts in a network. The communications that break the CTS rules will be obfuscated using CTO that implements obfuscation mechanisms. We do not obfuscate all network traffic because it will dramatically degrade network performance. In CHOAS, the network traffic will be first sent to IDS,such as Bro. If IDS judges that the traffic is suspicious, CTO module will obfuscate them through installing new flows into OpenFlow switches or modifying flows. Otherwise if the traffic is judged normal, it will be redirected to our Chaos Tower Structure module. The reasons for doing this are that adversaries often can bypass IDS through some unknown vulnerability attacks. CTS judges the risk of flows and divide them into expected connections and unexpected connections, detailed in section 3.2.1. Expected connections will be allowed. The unexpected connections will be obfuscated by Obfuscation module (CTO) according to different obfuscation levels.

\textbf{Chaos Tower Structure (CTS).} It is the module we design in the system to determine the communication rules. CTS builds a host hierarchy according to security level of information assets. The tower consists of several layers. Generally, important workgroups are placed in higher layers, whereas unimportant workgroups are placed in lower layers. The importance of every single node which can correspond to a host as well as the host cluster, is determined based on the importance degree of services and the vulnerability assessment score in the node. Then we build our model to control network traffic by defining which pairs of hosts can communicate in our topology. Further, according to the tower, the system divides connections into two types: expected and unexpected connections. 

\textbf{Chaos Tower Obfuscation (CTO). }It works on the basis of the CTS. It will obfuscate the suspicious traffic and unexpected traffic by a corresponding obfuscation level. Generally, traffics captured by IDS are subdivided into three levels. Then CTO obfuscates the connection according to the level.

We next elaborate the major processes of the whole system as shown in Figure 1. If an attacker tries to launch a request from a workgroup in relatively lower layers to a workgroup in higher layers, as indicated  by A and B in Figure 1, the system examines the corresponding connection. If it is found to be an abnormal connection, the system chooses an appropriate level to obfuscate the connection.

\subsection{Chaos Tower Structure and its Workflow}
The CTS is a combination of a tree structure and an oriented graph structure. We use a multi-branch tree in which to store the workgroup (a host is assigned to a specific workgroup according to its function or importance degree) and the tree defines the privilege of every workgroup. This ensures that most of the layer-jumping behavior is obfuscated. Nonetheless, some layer-jumping behavior is necessary (e.g., the two-way communication between a web server and a database server is necessary, although they are in distinct workgroups). We can define or modify the information conveniently by editing the ``Chaos Tower configure file'' in the controller. The tower structure with its strict hierarchy enables a more secure and more reliable network.

\subsubsection{Tower Construction}
In CHAOS, every host or subnet group will be examed and thus a corresponding risk level will be calculated. Risk levels are based on the underlying security metrics. In our system, we use the base score of Common Vulnerability Scoring System (CVSS)~\cite{29} to determine the intrinsic qualities of a vulnerability. CVSS base score includes two factors, \textit{exploitability of vulnerability} and \textit{impact of vulnerability}. CVSS classifies all the vulnerabilities depending on their features and effects and thus concludes several different kinds of vulnerabilities, such as SQL injection and Buffer overflow. For all these kinds of vulnerabilities, CVSS assigns different score to signal the importance of the vulnerability. And in addition to CVSS score, another critical factor is service importance value (SIV). Normally, some hosts are more valuable than others. Thus, we adopt service importance value to represent service's inherent value. It's worth mentioning that in different networks, the same service may be valued different. That's the reason why we set the SIV table as a part of configuration that administrators should define before the system works. In our system, we introduce the following generic equation to incorporate the CVSS base score and service importance value:
\begin{displaymath}
RL(h)=\sum_{v\in V(h)} (\alpha \times SIV(s) + (1-\alpha)\times CVSS(v))
\end{displaymath}
Where $RL(h)$ is the risk level of node $h$; $V(h)$ is a function to return all vulnerability contained in thehost $h$; $SIV(h)$ is a function to return the service importance value of the service s; and $CVSS(v)$ is a function to return the CVSS base score of the vulnerability $v$. We also introduce the weight coefficient \begin{math} \alpha(0\le\alpha\le1) \end{math} that allows an administrator to determine how important the service is. The value \begin{math} \alpha \end{math} can be increased, in which case the service is more important. Otherwise, the administrator can decrease the value of \begin{math} \alpha \end{math} to weaken the influence of the service but emphasize the influence of the possibility that the hosts would be attacked. According to this given information, we can continue building the original tower, which contains several layers. Each of these layers contains several workgroups, each of which includes several hosts that provide similar functions. CTS also can use some weights such as time, to further define access rules. For example, some access requests can be only allowed in some periods.

After the risk level of each hosts or groups is calculated, we put them into different layers of CHAOS tower. Hosts in the same layers should have the same risk level. Layers with higher risk level will have higher position (e.g., BYOD (Bring Your Own Device)). To deal with the situation that many new devices might well be added to specific subnets, we further divide hosts in the same layer into serveral groups. Each group contains at least one host. The group division is dependent on the hosts distribution in physical networks. Hence, when there are new devices added to the tower. CHAOS first exams whether they can belong to one existed group or not, if not, its risk level will be calculated and thus it will be mapped to a new group in the corresponding layers.

In CHAOS, we deem that the more important and risky the host is, the higher the layer it is assigned to. These wgroups share some common traits, for example, they may be used to store some important network resources. In our system, the administrators can define those important hosts and specify their order of privilege by the risk level of group.

\begin{figure*}
\centering
\includegraphics[width=5.5in]{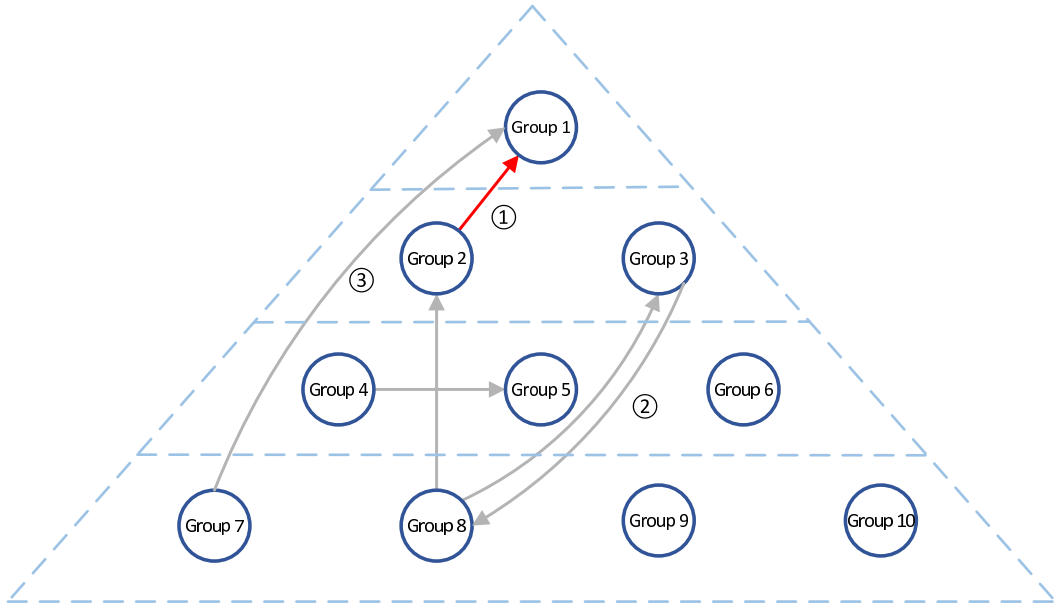}
\caption{Logical structure of the proposed CHAOS system. Red lines like \textcircled{1} represent the unexpected connections; gray lines from upper layers towards lower layers like \textcircled{2} represent the normal connections; gray lines from lower layers towards upper layers like \textcircled{3} represent special connections.}
\end{figure*}


\textbf{Expected Connections.} Expected connections include normal connections and special connections.

\begin{itemize}
\item{\textbf{Normal Connections.} Normal Connections, which represent the connections that are judged as normal connections by IDS. These connections correspond to normal services in intranet. It is worth mentioning that in each connection only one of the hosts is able to launch the connection while the other one is unable to do so. These behaviors can be regarded as harmless to properties of a network. Generally, normal connections are those connections that are allowed by the IDS.}

\item{\textbf{Special Connections.} In order to ensure availability of intranet services, even though some connections that a host belonging to higher layer accesses to a host belonging to lower layers are judged as suspicious flow by IDS, we still think it is expected connections. We release it temporarily, and record in the log so that administrator can analyze. }
\end{itemize}

\textbf{Unexpected Connections.}We define Unexpected Connections as those connections that are not included in the list of expected connections. Generally, these connections are not defined as being allowed and will be detected by our CHAOS system.

Here we consider an example to illustrate our proposed CHAOS system in more detail.  In Figure 2, Group 1 is placed to the top of tower due to its highest risk level. In this network, both Line 2 and Line 3 are regarded as expected connections (Note that 3 is special connection). And Line 1 is a unexpected connection even though it just transgresses only one layer.

\subsubsection{Exploiting the Tower}
In this section, we elaborate on how the system exploits the information in the tower. The system reacts differently for Expected and Unexpected Connections.

\textbf{Expected Connections.} We consider Expected Connections to be legal; thus, the system does not interfere with these connections.

\textbf{Unexpected Connections.} Attention should be paid to these connections. If confronted with an Unexpected Connection, the corresponding switch sends a packet-in message to the controller to allow it to decide how to proceed. If the connection extends from a higher to a lower layer, it is allowed. This kind of connection does not need to be obfuscated: should an adversary gain control of a host in the relatively higher layers, then a compromise of the lower layers would not be regarded as a great loss for the enterprise provided the CHAOS Tower is well defined by the administrator. On the contrary, if the connection is established by layer-jumping or occurs within a layer, it is considered abnormal and will be obfuscated.

\section{Obfuscation}
The steps listed above output unexpected connections from the CTS. 
 In our system, we implement three kind of obfuscations, which are host mutation obfuscation, port obfuscation and obfuscation based on decoy servers. The former in the three is for all connections, and the other two are for unexpected connections. We firstly introduce a parameter which enables the administrator of the network flexibly adjust the degree of obfuscation in the whole network so as to implement a differential obfuscation.Then we further discuss how CHAOS grades unexpected connections and apply corresponding obfuscations according to their degree of abnormality.
\textbf{CHAOS Parameter Assignment. }
We provide a parameter to administrators to enable  a flexible way to decide which hosts will be obfuscated while CHAOS takes risk level into consideration.  We incorporate this  threshold parameter \begin{math} T(0\le T \le 1) \end{math} to determine the obfuscation index. The smaller T is, the more traffic in the network will be obfuscated. For example, the parameter which is 0.5 leads to 50\% unexpected connections will be obfuscated. The initialization details are as follows.

\begin{enumerate}
\item Set the weight coefficient \begin{math} \alpha \end{math} according to different host.
\item Compute risk level of each host.
\item Assign each host to the corresponding layer according to the risk level.
\item Divide the remaining hosts into several workgroups according to the network structure (i.e., hosts connected to the same switch are assigned to one workgroup).
\item Set a threshold factor T.
\end{enumerate}

\textbf{Host mutation obfuscation. } This obfuscation method is proposed in \cite{7}. The mechanism is shown in the right-hand side of Figure 3. The OpenFlow controller frequently assigns a random virtual IP (vIP) to each real IP (vIP). When \emph{Host1} initiates the connection to \emph{Host2} and sends a initial packet using real source IP (\emph{r1}) and real destination IP (\emph{r2}), the first OF Switch that captures the initial packet (\emph{OF Switch 1}) encapsulates and sends the packet to SDN controller, where a rIP-vIP mapping table is stored, and maps \emph{r1} and \emph{r2} to corresponding virtual IPs (\emph{v1} and \emph{v2}). When the initial packet reaches the OF Switch that is nearest to \emph{Host2} (\emph{OF Switch n}), a similar reverse mapping is executed, changing vIPs back to rIPs, namely, \emph{v1} to \emph{r1} and \emph{v2} to \emph{r2}. In this sense, packets in the middle (between \emph{OF Switch 1} and \emph{OF Switch n}) only contain virtual IPs so that conceal real host IPs.

\begin{figure*}
\centering
\includegraphics[width=6.5in]{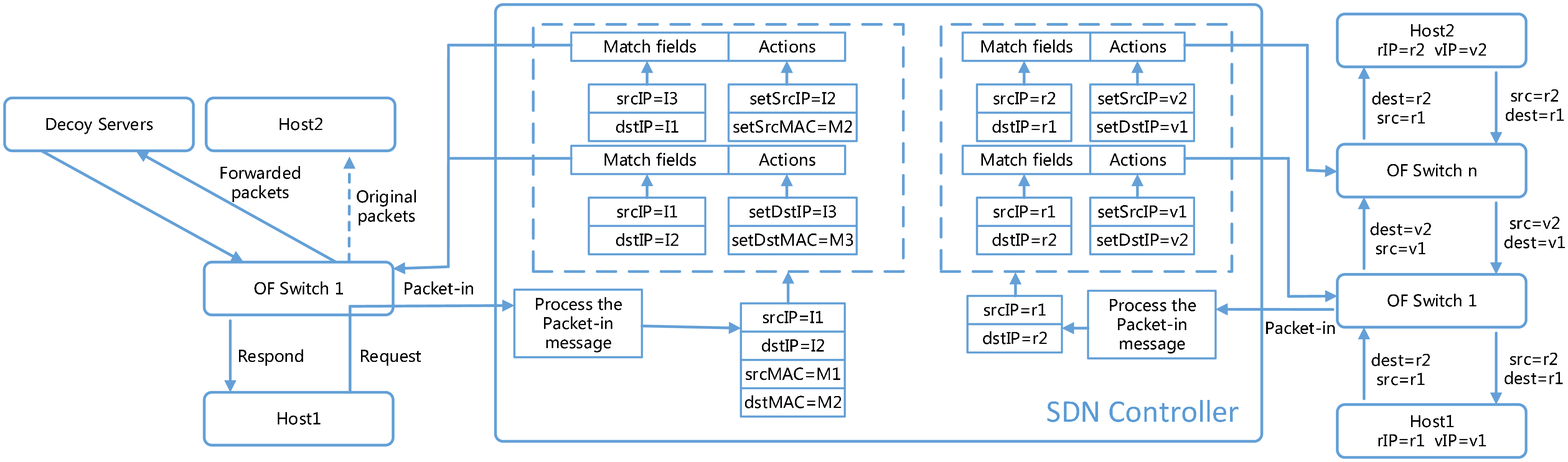}
\caption{Mechanism of host mutation and decoy-servers-based obfuscation}
\end{figure*}

\textbf{Port obfuscation. }This technique is aimed to defend port-scanning-based attack. In this case we inject some entirely fake information into responses as well as hiding some real information.  As is shown in Figure 4, when IDS detects a port scanning, CHAOS system will inject fake packets into the real packets by generating corresponding acknowledgement to obfuscates the result of the port scanning. For instance, when a TCP scan is detected and port obfuscation is applied, the TCP packets will be fetched by switch and sent to the controller through Packet-in. Then the controller will analyze the packet, generate a corresponding Packet-out and send it to the switch. The acknowledgements of some injected packets are 0 while some are 1.  Whether to inject or modify the packets is generally on a random basis. Therefore, the results of port scanning will show a certain degree of randomness and fuzziness.

\begin{figure}
\centering
\includegraphics[width=5in]{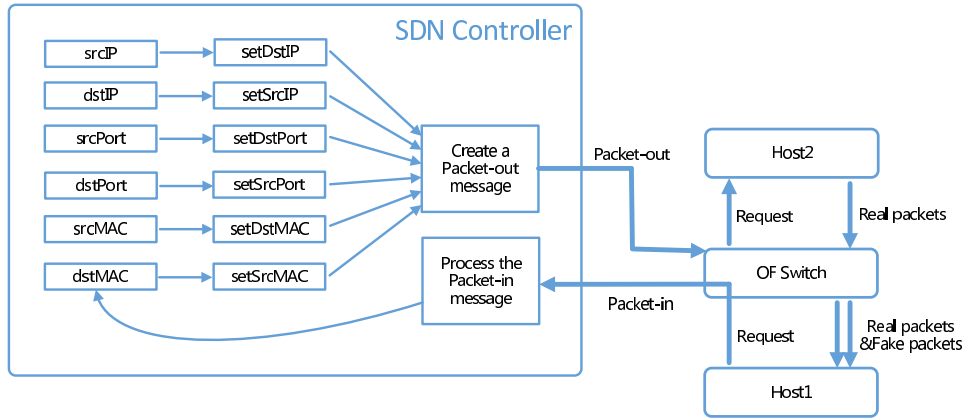}
\caption{Mechanism of port obfuscation}
\end{figure}

\textbf{Obfuscation based on decoy servers.} In CHAOS system, we deploy a number of decoy servers. In this case, our system will forward the unexpected connections to the decoy servers. As is shown in Figure 3, when a host launch a request, our system can analyze the packets and install flows into the switch, which will forward the unexpected connections to our decoy servers. In this way, suspicious users can only access various decoy servers. The services we deployed in the decoy servers can further help we discover the real attackers.

These three obfuscation strategies are applied under different circumstances. In the tower, we use the threshold factor to determine which strategy is applied. As showed in Table 2, the host mutation obfuscation is applied to all connections. The port obfuscation and the obfuscation based on decoy servers are applied to unexpected connections, but which one is selected depending on the threshold factor. The risk value is determined by calculating the ratio of leapfrog access number to the total number of the layers.

\begin{table} \scriptsize
\centering
\caption{Obfuscation strategy}
\begin{tabular}{|l|c|} \hline
Scope&Obfuscation method\\ \hline
All connections & Host mutation obfuscation\\ \hline
Unexpected connections(risk$\le$threshhod) & Port obfuscation\\ \hline
Unexpected connections(risk$>$threshhod)& Obfuscation based on decoy servers\\ \hline
\end{tabular}
\end{table}

In addition, we introduce a parameter named RandomIndex (0$<$RandomIndex$<$1) to define the possibility of CHAOS performing obfuscation, i.e., the closer RandomIndex is to 0, the higher the likelihood of CHAOS injecting fake information into the network. Here we use a pseudocode to clarify the process. We define srcLayer as the layer in which the host launches the request and dstLayer as the layer in which the host responds. Then we define altitude as the difference in height between these two respective layers (i.e., the height of scrLayer minus the height of dstLayer).

Our implementation of obfuscation contains two aspects.
First, as most network mapping tools perform their operations by using ICMP packets and TCP or UDP scans. ICMP messages are typically used to verify connectivity or reachability of potential targets. TCP and UDP port scans are used to identify running services of a target. Replies (TCP RST, silent drop or ICMP unreachable) to scans can also reveal what services are allowed or filtered through transit devices. Additionally, the TTL field of IP packets is used to identify the hop distance between the target and the destination. SDN-enabled devices can be used to confuse the reconnaissance. For example, traffic to a destination that can be blocked according to a filtering policy can be silently dropped and SDN utilities can generate varying responses that will confuse the attacker. In the case of traffic that is permitted by the filtering policy (that is, it is legitimate), the SDN policy does not interfere. The action for each packet is kept in a buffer to ensure consistent behavior. As a result of this algorithm, random ports will appear to the scanner as being open. Digging deeper in order to identify services running on these fake open ports would require more resources from the attacker. \cite{24} Secondly, the controller determines the type of connection (i.e., via srcIP or dstIP) and installs necessary flows in all OF-switches in the path. These flows will change the srcIP and dstIP of each packet (assuming srcIP changed to be vsrcIP and dstIP change to be vdstIP) so that the packet will be different from what they are actually. But meanwhile, these flows will also make sure that the packet can be sent to the destination host by changing the vsrcIP and vdstIP to srcIP and dstIP in the end. Each connection must be associated with a unique flow, because the rIP-vIP translation changes for each connection. This property guarantees the end-to-end reachability of hosts, because the rIP-vIP translation for a specific connection remains unchanged regardless of subsequent mutations \cite{7}.

The process is presented as Algorithm 1. Firstly, if we find that the Packet-In message comes from the source switch or destination switch of the packet, we will install flow tables of host mutation. Then, the connection will be judged to be a expected connection or an unexpected connection determined by its altitude between its source layer and its destination layer. For expected connections, the packet will be forward directly. But for unexpected connections, the packet will be obfuscate or forwarded to a decoy server according to the RandomIndex shown before or be dropped directly if the altitude is bigger than the threshold configured by administrator.

\begin{algorithm}
\floatname{algorithm}{Algorithm}
\footnotesize
\caption{CHAOS}
\label{alg1}
\begin{algorithmic}
   \REQUIRE $packetIn p,  Inf, Sup, RandomIndex$;
   \COMMENT{HEIGHT is the height of the tower}
     \IF{$isFromSrcSwitch(p) or isFromDstSwitch(p)$}
        \STATE $installHostMutationFlows(p)$;
     \ENDIF
    \STATE $srcLayer \gets getSrcLayer(p)$;
    \STATE $dstLayer \gets getDstLayer(p)$;
    \STATE \(\Delta\)$Altitude \gets srcLayer-dstLayer$;
    \STATE $Possibility\gets random[0,1]$;
    \IF{ \(\Delta\)$Altitude \geq 0$}
        \STATE $Forward(p)$;
    \ELSE
        \STATE \(\Delta\)$Altitude \gets -$\(\Delta\)$Altitude$;
        \IF{ \(\Delta\)$Altitude/HEIGHT \leq threshold$}
            \IF{$isRequestPacket(p) \textbf and Possibility \geq RandomIndex$}
                \STATE $PacketOut(p)$;
            \ELSE
                \STATE $ForwardToDecoyServer(p)$;
            \ENDIF
        \ELSE
             \STATE $installForwardingFlows(p)$;
        \ENDIF
    \ENDIF
\end{algorithmic}
\end{algorithm}

\section{Implementation and Evaluation}

\subsection{System Implementation}
The structure of our system is shown in the Figure 5. The routing was managed entirely by the Floodlight controller and monitored by Bro. We implemented three modules. The first one we implemented is the CHAOS tower module, the purpose of which is to build the CHAOS tower and get suspicious flows from Bro automatically. Then, we implemented the Obfuscation module in Floodlight, which is supposed to implement obfuscation on those unexpected connections. Finally, we implemented the CHAOS management module which allows administrators to further configure their networks. When the system starts, an initialization process occurs  to obtain configuration provided by administrators and to pass the information to the corresponding modules such that those modules are able to work well later.

\begin{figure}
\centering
\includegraphics[width=5.5in]{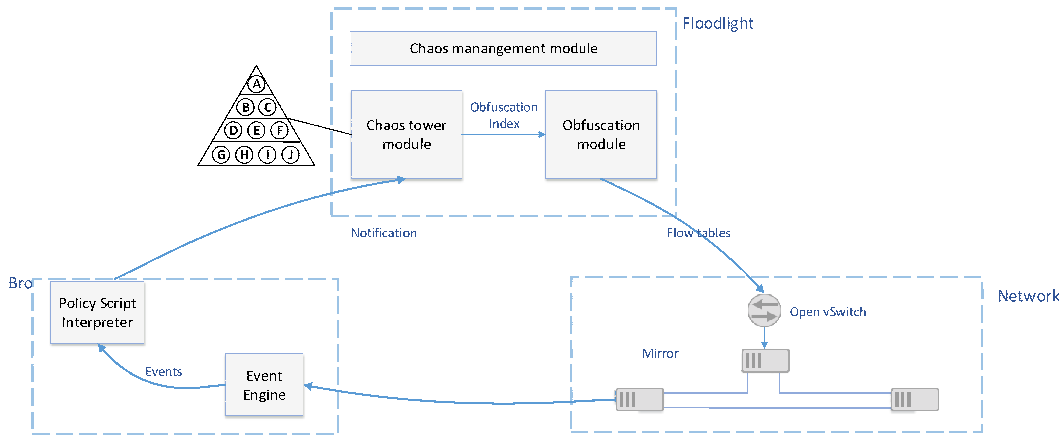}
\caption{System implementation}
\end{figure}

We provide an implementation of obfuscation with Bro's warning message. In the beginning, we push flow tables into switches so that all flows are allowed. Then, we use Bro to monitor the network. When suspicious flows detected, the tower will determine the corresponding obfuscation index and transfer it to obfuscation module. After that, corresponding flow tables will be updated to make sure that the obfuscation works in the network.

Finally, we claim that our Obfuscation rules only drop responses partially. We only drop those parts containing sensitive information from the target host to prevent the adversary from accessing all the information from the target host.

\subsection{Scanning and Footprinting Test}
Footprinting and scanning are techniques for gathering information about computer systems in networks. These techniques are implemented by various security auditing tools as the first step when launching an attack. Nmap \cite{8} and the scanner modules in Metasploit \cite{25} contain many payloads to gather sensitive information from target machines, whereas Nessus \cite{26} and WVS \cite{27} focus on vulnerability detection and exploitation.

In our test, we used Nmap to evaluate the information obfuscation ability of CHAOS. Nmap uses raw IP packets in novel ways to determine which hosts are available on the network, which services (application name and version) those hosts are offering and which operating systems (and OS versions) they are running, which type of packet filters/firewalls are in use, and many other characteristics \cite{8}. Our test involved configuring some vulnerable hosts in the network, after which we used Nessus to detect vulnerabilities to test whether CHAOS would be able to confuse and deceit Nessus.

We tested the performance of our system by launching a series of attacks under different circumstances. We consider three situations against Nmap. In the first, the network was unprotected; in the second, moving target defense protection was implemented with a random algorithm obtained from \cite{24}; and in the third, our CHAOS system was implemented. When simulating the attack, we used Nmap to scan the entire network several times. Based on its response and the reality of its given circumstances, we concluded the result (Figure 4 and Figure 5). Besides this, we used a ping command to test the effect of our system on normal traffic (Figure 6).

\subsection{Results}

We carried out our experiments in CloudLab \cite{25} and deployed the network shown in Figure 2.

First, we used Nmap to determine whether our CHAOS system was able to deceit the security tool. There are two situations involved in this experiment. We selected the hosts of the Group 4 and the Group 3 in Figure 2; thus, the obfuscation index is 0.5, so obfuscation based on decoy servers will work then.

We define information disclosure percentage(IDP) as our index and calculate it by following formulas. ID is the amount of information the adversary fetches from the victim.

 \begin{displaymath}
 IDP_{CHAOS}=ID_{CHAOS}/ID_{NONE}\end{displaymath}
 \begin{displaymath}
 IDP_{MTD}=ID_{MTD}/ID_{NONE}\end{displaymath}

Figure 6(a) shows the percentage of information disclosure of an unprotected network and a network (Level 2) protected by CHAOS as a function of the number of times the network was scanned by Nmap. The figure shows that, for the network protected by CHAOS, the percentage of information disclosure is decreased effectively.

\begin{figure*} [t]
\centering
\subfigure[Information disclosure]{
\begin{minipage}[b]{4.5cm}
\includegraphics[width=1\textwidth]{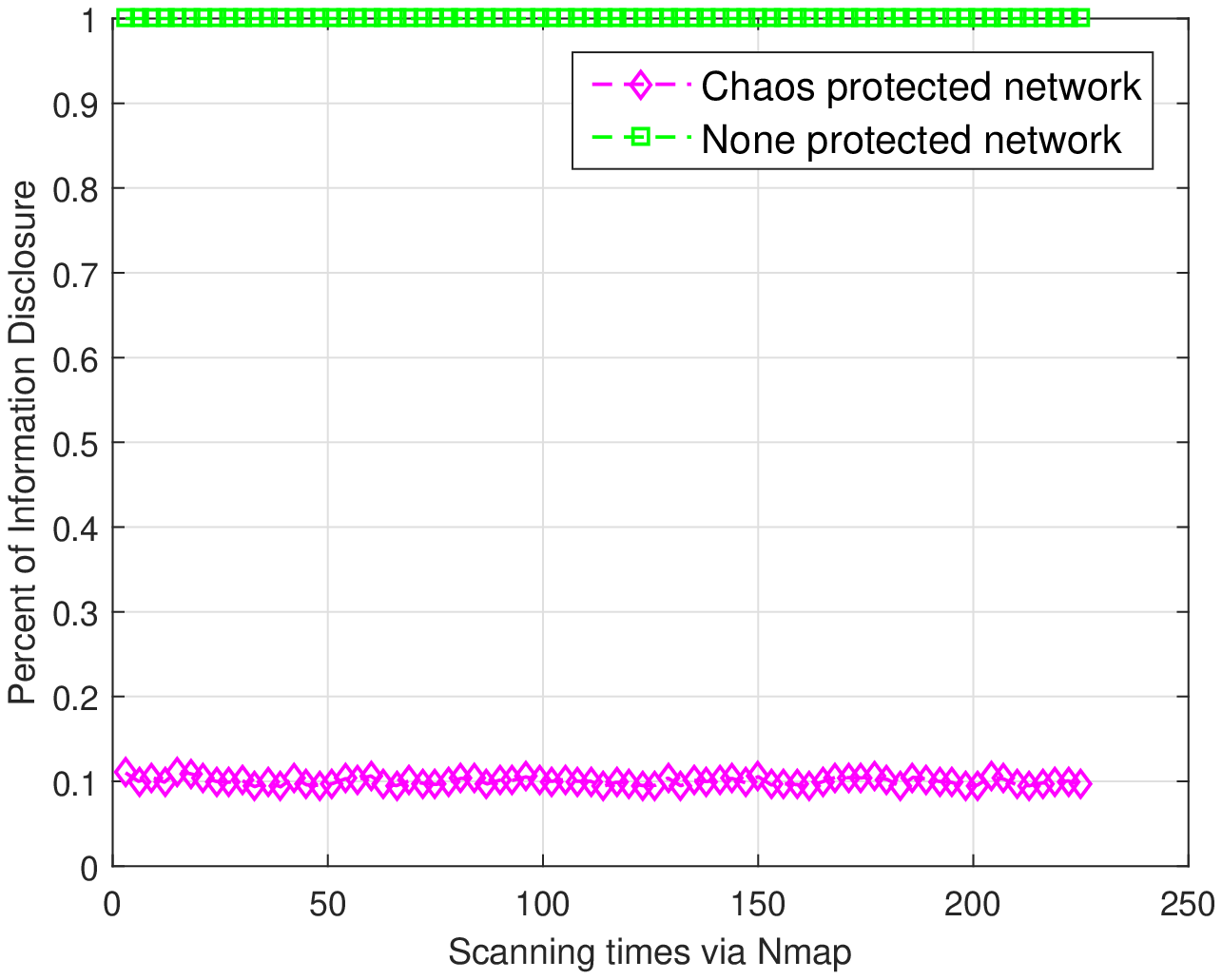}
\end{minipage}
}
\subfigure[Information disclosure with respect to threat degree]{
\begin{minipage}[b]{4.5cm}
\includegraphics[width=1\textwidth]{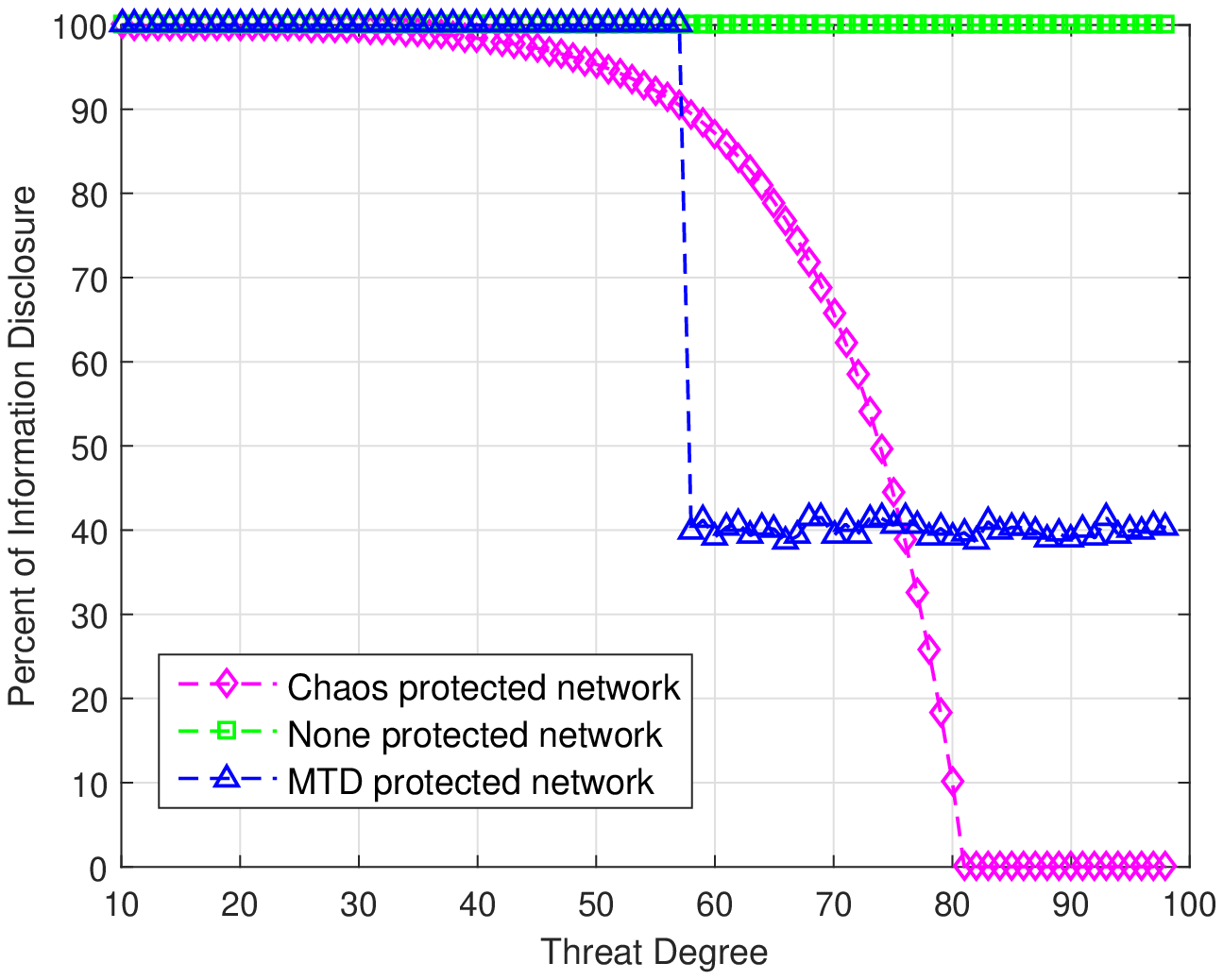}
\end{minipage}
}
\subfigure[Delay time with respect to packet count]{
\begin{minipage}[b]{4.5cm}
\includegraphics[width=1\textwidth]{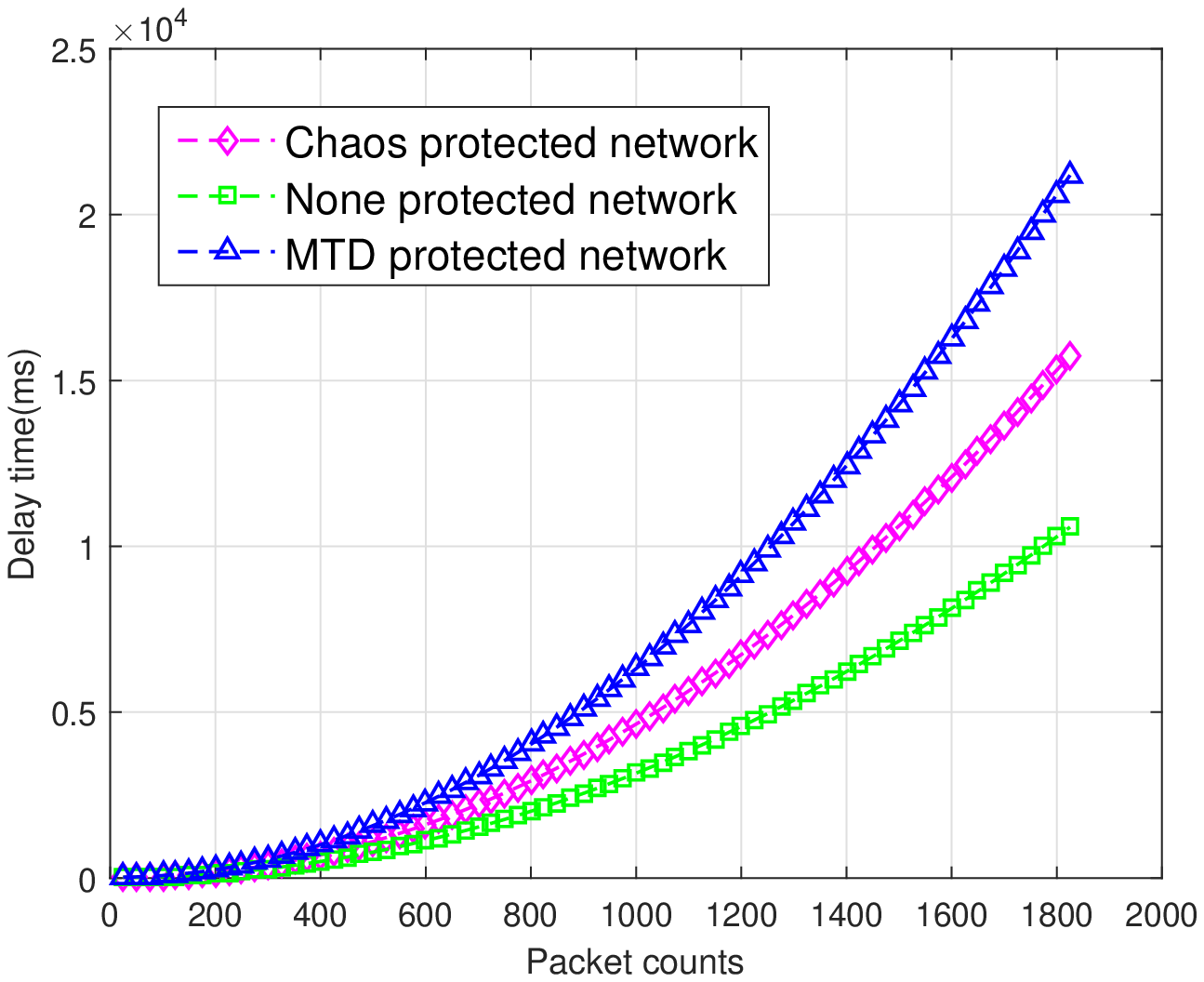}
\end{minipage}
}
 \caption{Test results} \label{fig:1}
\end{figure*}

Secondly, we studied the correlation between the degree of threat of the adversary and the information disclosure he would experience. Figure 6(b) shows the information disclosure in an unprotected network, a network protected by CHAOS, and an MTD protected network \cite{25}, all of which face different degrees of threats. The MTD protected network obfuscates all the packets by some static policies. Thus it is able to decrease information disclosure when the threat reaches a certain degree, but does not decrease information disclosure further when the degree of threat is elevated beyond that certain degree, because of its static solution. However, the network protected by CHAOS decreases information disclosure when the degree of threat is elevated. There is almost no information disclosure when the threat reached a very high degree.

After that, we compared the performance cost of the three networks. As above, we compare the network protected by CHAOS with the unprotected and MTD protected networks. We use the example shown above to test the performance of these systems and to measure the average delay time of the connections under each system. Figure 6(c) shows the delay time of the unprotected network, the network protected by CHAOS, and the MTD protected network with changing package counts. We conclude that both the networks protected by CHAOS and MTD increase the delay time to some extent, although the network protected by CHAOS has a reduced delay compared to that protected by MTD. Thus, our system enables the network to perform faster. We discovered that the transforming speed of our system is faster than that of random obfuscation system especially when the network is crowded.

The result above can be understood in terms of the following factors.

First, we use Bro to monitor the network and transfer those suspicious flows. The important point is that Bro runs stand-alone so it makes quite few effects to the speed of the network.

Then, the CHAOS Tower is also a factor that reduce the delay time. We assume that the CHAOS Tower is to be built as a binary tree in the network and the number of layers is L; hence,
\begin{displaymath}
N=2^L-1
\end{displaymath}
We consider a situation in which each workgroup sends a request to the remaining groups, which means that the sum of the connections the unprotected situation and the MTD solution would have to process would be:
 \begin{displaymath}
 C_{NONE}=0\end{displaymath}
  \begin{displaymath}C_{MTD}=N*(N-1)\end{displaymath}
 However, we only need to obfuscate the connections from the lower layers toward the higher layers in our CHAOS system, the number of which is:
 \begin{displaymath}C_{CHAOS}=\sum_{i=1}^{L-1} (2^i*(2^i-1))\end{displaymath}

In the end, we launched several real attacks to testify robustness of our system. We employ some vulnerable hosts in the network. In the experiment, MS 08-067 is the vulnerability that we test. The hosts can be easily attacked by any pentesting tools which contains payload of MS 08-067. Actually, in Chaos Tower, we employ a vulnerable host in each layer. Then we use one of them to play the role of attacker in turn. Figure 7 shows the results of the unprotected network, the network protected by IDS with CHAOS, and the network protected only by IDS. We conclude that in the network protected by IDS with CHAOS, only a few attacks directed to hosts belonging to adjacent layers succeeded. However, in the network protected only by IDS, most attacks succeeded in the end. The worst is in the unprotected network nearly all attacks succeeded. Thus, our system can decrease the success rate of such kind of attacks significantly.

\begin{figure}
\centering
\includegraphics[width=4in]{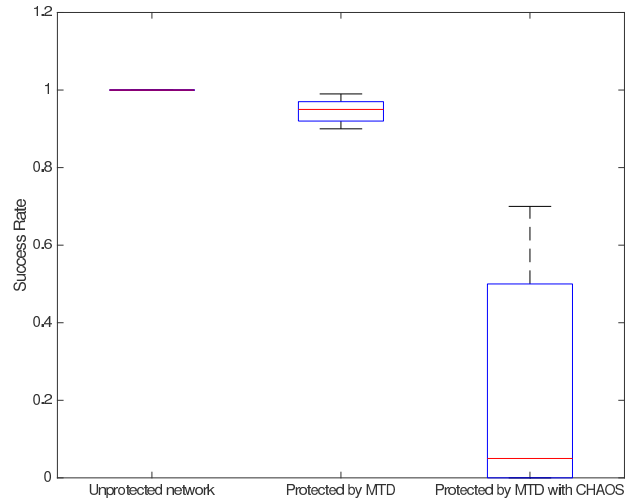}
\caption{Attack testing}
\end{figure}

\section{Related Work}
Several researchers have reported work on MTD. Kewley et al. \cite{15}  performed the initial research in the area of dynamic network defense and proved that dynamic network reconfiguration, such as randomly changing the IP address and port numbers, would effectively inhibit an adversary's ability to gather intelligence, and thus degrade the ability to successfully launch an attack. Al-Shaer proposed MUTE, a moving target defense architecture \cite{12}, which implements the moving target through random address hopping and random finger printing. Furthermore, they presented BDD, a model for creating a valid mutation of network configuration. Zhang et al. \cite{11} investigated the application of moving target defenses to network security and presented a high-level architecture of the MTD system. Their simulation results show the potential for MTD to be effective in preventing attacks against computer networks. Furthermore, they proposed a formal theory to describe the MTD system and its basic properties and formalized the MTD entropy hypothesis, which states that the greater the entropy of the system configuration, the more effective the MTD system \cite{13} \cite{10}.  Stallings proposed the use of SDN in the implementation of MTD mitigations. Jafarian \cite{7} proposed OpenFlow Random Host Mutation (OF-RHM), which uses OpenFlow to develop an MTD architecture that transparently mutates host IP addresses with high unpredictability, while maintaining configuration integrity and minimizing operational overhead.

However, current network-based MTD obfuscates networks indiscriminately that makes some networks services unavailable, e.g. some key services like web and DNS because some information of these services have to be opened to the outside and remain real. If MTD obfuscates these services fully, it will return users with virtual IPs and ports, making these services unable to use. Moreover obfuscation will affect the performance of networks. To obfuscate hosts indiscriminately will severely reduce the performance of networks undoubtedly. In contrast to above work, CHAOS discriminately obfuscates hosts with different security levels in networks.

Zhang \cite{14} proposed to construct an incentive-compatible moving target defense by  periodically migrating virtual machines (VMs), thereby making it much harder for adversaries to locate the target VMs. Al-Shaer \cite{16} proposed to defend against DDoS attacks by  migrating  virtual networks (VNs) to dynamically reallocate network resources. Different from their work, CHOAS leverage SDN features to obfuscate network information instead of migrating target objects.

Previous research involving memory address space randomization \cite{19} \cite{18} \cite{20}, instruction set randomization \cite{21}, and software diversification \cite{22} \cite{23}, also used the idea of a moving target to increase the attack difficulty and cost by enlarging the exploration surface or moving the attack surface.  The objective of our work is to enhance network security; hence, the aspects mentioned here are not discussed in detail.

\section{Conclusion}
MTD is able to create a type of changing network so as to increase the difficulty and cost for an adversary aiming to launch a network attack. However, MTD may disrupt the flow of normal traffic when performing obfuscation. This paper proposes an SDN-based MTD system named CHAOS which discriminately obfuscates hosts with different security levels in networks so as to keep some key services available and low performance cost. CHAOS incorporates the Chaos Tower Structure to represent a hierarchy of all the hosts on the network and leverages SDN features to obfuscate the attack surface to enhance the unpredictability of the networking environment. CHAOS offers rapid obfuscation of unexpected network traffic, but does not interfere with normal traffic. The evaluation shows that a network protected by CHAOS can effectively lower the percentage of information that is disclosed.

\Acknowledgements{This work is sponsored by the National Natural Science Foundation of China granted No. 61402342, 61173138 ,61103628 and National Basic Research Program of China (973 Program) granted No.2014CB340600.}


\end{document}